\documentclass[runningheads,citeauthoryear]{apinv}
\usepackage{epsfig,cite,graphics}

\usepackage[T2A]{fontenc}
\usepackage[cp1251]{inputenc}
\usepackage[bulgarian,english]{babel}

\def\Ha {H$_\alpha$\,}
\def\Msun {$M_\odot$\,}
\def\Zsun {$Z_\odot$\,}

\begin{document}
\title{The VLT-FLAMES Tarantula Survey}
\titlerunning{The FLAMES Tarantula survey}
\author{N.~Markova\inst{1}, C.~J.~Evans\inst{2}, N.~Bastian, Y.~Beletsky, 
J.~Bestenlehner, I.~Brott, M.~Cantiello, G.~Carraro, J.~S.~Clark, P.~A.~Crowther, 
A.~de~Koter, S.~E.~de~Mink, E.~Doran, P.~L.~Dufton,
P.~Dunstall, M.~Gieles, G.~Gr\"{a}efener, V.~Henault-Brunet, 
A.~Herrero, I.~D.~Howarth, N.~Langer, D.~J.~Lennon, J.~Ma\'iz~Apell\'aniz, 
F.~Najarro, J.~Puls, H.~Sana, S.~Sim\'on-D\'iaz, S.~J.~Smartt, V.~E.~Stroud, 
W.~D.~Taylor, J.~Th.~van~Loon, J.~S.~Vink, N.~R.~Walborn}
\authorrunning{Markova et al.}
\institute{ 
\inst{1}Institute of Astronomy with NAO, BAS  
\newline
\inst{2}{UK Astronomy Technology Centre, Royal Observatory Edinburgh}
\newline
}
\email{nmarkova@astro.bas.bg}

\papertype{contribution}
\maketitle
\begin{abstract}
The Tarantula survey is an ESO Large Programme which has obtained multi-epochs spectroscopy of over 800 massive stars in the 30 Dor region in the Large Magelanic Cloud. Here we briefly describe the main drivers of the survey and the observational material derived. 
\end{abstract}


\section{Introduction}

Despite their scarcity, massive O-type stars and their descendants
play an important role in the history of the Universe (from the
earliest times to the present day). They are the main engines which
drive the chemical and dynamical evolution of galaxies, enriching the
interstellar medium with heavy elements, creating H\,{\sc ii} regions
and exploding as supernovae. In the distant Universe, they dominate
the integrated UV radiation in young galaxies. Massive stars are
potentially also key objects for studying and understanding phenomena
such as cosmic re-ionisation and $\gamma$-ray bursts.

The VLT-FLAMES survey of Massive stars (FSMS, Evans et al. \cite{evans05}, Evans et al. \cite{evans06} was a European Southern Observatory (ESO)
Large Programme which obtained high-resolution spectroscopy of about
700 massive stars from seven cluster fields in the Milky Way and
Magellanic Clouds, with the primary objective to study the effects of
metallicity on stellar evolution.  Some of the key results include: a
precise quantification of the theoretically predicted metallicity
dependence of the winds from O-type stars (Mokiem et al. \cite{mokiem07}); determinations of chemical abundances in the Clouds (Hunter et al. \cite{hunter07}, Trundle et al.\cite{trundle}); first evidence of a 
metallicity
dependence in the effective temperature of B-type stars at a given spectral
type (Trundle et al. \cite{trundle}); new insights into the role of rotational
mixing in the atmosphere of B-stars (Hunter et al.  \cite{hunter08b},
\cite{hunter09}); and a quantitative verification of the assumption that low
metallicity massive stars should rotate faster (Hunter et al.
\cite{hunter08a}).  (A summary of the results from the FSMS can be 
found in Evans et al. \cite{evans08})

Building upon the experience gained through the FSMS, the FLAMES
consortium initiated a new multi-epoch spectroscopic survey of the
Tarantula Nebula in the Large Magellanic Cloud -- the VLT-FLAMES
Tarantula Survey (P.I. Evans).  The Tarantula Nebula (also known as
30~Doradus) is an immense star-forming region at the eastern end of
the stellar bar in the LMC, and the most active starburst region in
the Local Group.  A significant amount of the total energy emitted
by the nebula comes from R136, a compact cluster located at its core,
thought to host some of the most massive stars known (Crowther et
al. \cite{crowther}). Due to its proximity, its relatively low 
reddening, and the
favorable inclination of the LMC, the Tarantula Nebula is one of the
best laboratories to study the formation and evolution of massive
stars.

\section{Project aims}

A detailed description of the scientific objectives of the project
will be provided in the survey overview paper (Evans et al.  in
preparation); here we briefly comment on three of the main drivers.

\paragraph {\it Predictions of evolutionary models:} 
Evolutionary models incorporating stellar rotation (Meynet and Maeder
\cite{MM}2000, Heger and Langer \cite{HL}) predict significant 
enhancement of the
surface helium and nitrogen abundance in massive O-type stars due to
the effects of rotational mixing.  The models also predict that stars
with initial masses larger than 15~\Msun, will either slow down (at
Z=\Zsun) or remain at almost constant rotation rates (at Z lower than
\Zsun) throughout their main-sequence (MS) lifetimes, suggesting that
massive O-type stars in the Magellanic Clouds should rotate faster
than their Galactic counterparts.  While some of these predictions
have been confirmed by observations, others are not (e.g., Penny and Gies 
\cite{PG} and references therein). The FLAMES Tarantula Survey aims
to investigate the effects of rotation at the high-mass end of the
Hertzsprung-Russell (H-R) diagram and, in particular, to test the
model predictions with regarding to rotational mixing of core-processed
material to the stellar surface.

\paragraph{\it Massive binaries:} If undetected, binarity
(multiplicity) can significantly modify the results of spectroscopic
and photometric analysis. The fraction of massive binaries in 30 Dor
is expected to be large (Bosch et al. \cite{bosch}), so to identify and
characterize the binaries among the FLAMES sample is a crucial task.

The multi-epoch observation strategy adopted for the survey allows us
to obtain clear signatures of short-period binarity in the majority of
the targets, while also providing information for some of the long-period
binaries as well, This will put robust constraints on the total binary
fraction in~30 Dor. Armed with these new data and the results from the
model atmosphere analysis we will then be able to test the model
predictions of both single-star and binary evolution.

\paragraph{\it Dynamical mass of R136:} 
The effect of binaries on the dynamical mass estimates of young
stellar clusters was highlighted recently by Gieles et al. \cite{gieles}.
The multi-epoch FLAMES observations will be used to obtain precise
radial velocities for our targets in and around R136. We will then
determine the velocity dispersion for single massive stars (i.e.
excluding the detected binaries) to derive a dynamical mass for the
cluster.  This will provide an important external estimate to the
debate regarding the photometric mass of R136, which hinges on the
low-mass form of the initial mass function (Sirianni et al. \cite{sirianni},
Andersen et al. \cite{andersen}).

\section{Observational material}

The Fibre Large Array Multi-Element Spectrograph, FLAMES
(Pasquini et al. \cite{pasquini}), is on UT2/Kueyen of the Very Large
Telescope (VLT) at Cerro Paranal. FLAMES can be used to feed two
optical spectrographs: the medium-high resolution spectrograph
Giraffe, and the high-resolution spectrograph UVES. The Giraffe
spectrograph works in combination with three different types of fibre
systems - Medusa (single fibres), deployable integral field units
(IFUs), and ARGUS (a monolithic central IFU); UVES is fed by only one
fibre system attached to its red arm.

The observational strategy of the new survey builds upon the
experience gained from the FSMS and includes multi-epoch observations
with the Giraffe spectrograph in its Medusa and ARGUS IFU modes, plus
using the feed to UVES.  The sample comprises $\sim$1\,000 stars
brighter than V\,=\,17 mag. (to ensure adequate signal-to-noise, S/N)
with no limitation in color (to avoid selection biases and achieve a
good representation of the upper part of the H-R diagram).  The
targets were primarily selected from unpublished imaging of 30~Dor,
distributed such they cover its full spatial extent and outwards into
the surrounding field stars and other nearby OB associations. The
total field spans a diameter of about 20~arcminutes from the center of
R136. Observational details are summarized in Table~\ref{observations}.
Individual comments on each observational component are outlined
below.

\paragraph{MEDUSA observations:} The primary dataset for our survey
was obtained with the Giraffe spectrograph in the Medusa mode. To
build-up the sample, a total of nine Medusa configurations were
used\footnote{Each Medusa configuration allows up to 132 targets
  (including sky fibres), distributed over a 25$'$ field-of-view to be
  observed simultaneously.} Each target was observed at three of the
standard Giraffe settings (LR02, LR03 and HR15N), providing a total
wavelength coverage from about 4\,000 to about 5\,000 \AA, plus the
region around \Ha needed to constrain stellar wind properties.

For detection of massive binaries, each of the Medusa targets was
re-observed at three additional epochs using the LR02 setting. The
execution of these was constrained such there was a minimum of 28 days
between the follow-up epochs.  To put constraints on long-period
systems, a final epoch was obtained one year later then the main
observations. The S/N ratio of the combined Medusa data is greater
than 50 for all of the O- and B-type stars.
\begin{table}[t]
\begin{center}
\caption{Summary of the spectral coverage and resolving power ($R$) of the
FLAMES Tarantula Survey observations.}\label{observations}
\tabcolsep2.0mm
\begin{tabular}{llll}
\hline
\\
\multicolumn{1}{c}{Mode}
&\multicolumn{1}{l}{Setting}
&\multicolumn{1}{r}{$\lambda$-coverage [\AA]}
&\multicolumn{1}{c}{$R$}
\\
\hline
\hline
Medusa & LR02  & 3960 to 4564 & 7\,000\\
Medusa & LR03  & 4499 to 5071 & 8\,500\\
Medusa & HR15N & 6442 to 6817 & 16\,000\\
ARGUS  & LR02  & 3980 to 4570 & 10\,500)\\
UVES   & Red arm & 4175 to 6200 & 47\,000\\
\hline
\end{tabular}
\end{center}
\end{table}
\paragraph{ARGUS IFU observations:} We also employed the Giraffe
spectrograph in its ARGUS IFU mode to sample the denser central region
in and around R136. Five ARGUS pointings were observed with the LR02
setting, see Table~\ref{observations}. To identify and characterize
massive binaries, each ARGUS pointing was observed at five different
epochs.  Following the Medusa strategy, the first two observations
were executed with no time restriction; the second and the third, and
third and fourth epochs were separated by a minimum of 28 days; the
final observations were obtained about one year later.

\paragraph{UVES observations:} In parallel to the ARGUS observations a
small sample of 25 stars were observed using the six fibres feeding
the red arm of UVES. Twenty of these were in common with the
Giraffe-Medusa sample; five are unique.

\paragraph{Complementary observations:}
To expand our view of the 30 Dor, the Tarantula Survey is supplemented
with data from two external sources.

\begin{itemize}
\item{VLT-SINFONI near-IR spectroscopy:} $K$-band spectroscopy was
obtained of the central arcminute around R136 using the SINFONI IFU
(P.I. Gr\"{a}fener). These data will be used to investigate the wind
properties and to constrain the clumping factor.
\item{Faulkes Telescope South observations:} As part of the Faulkes 
off-line queue and also the schools educational programme, there is an
on-going monitoring campaign using the Faulkes Telescope South to
obtain Bessel $B$ and $V$, SDSS $i'$, and Pan-STARRS $Y$
observations of many of our FLAMES targets.  This will enable more
detailed analysis of the binary systems identified via the multi-epoch
spectroscopy.
\end{itemize}

\section{Current status}

All the observations have been obtained, with a total of about 22\,000 
spectra collected. The data have been reduced and released to the 
consortium, with work now proceeding on multiple strands of research.
Based on the reduced spectra from the LR02 settings, and results from
a preliminary radial velocity analysis, rough classification of the
Medusa targets was performed. The sample comprises approximately 300
O-type stars, 500 B-type stars, 20 WR/slash stars, 90 stars with
cooler spectral types (but with radial velocities consistent with
their membership of the LMC) and about 100 foreground stars. As a
first step towards a more detailed classification, a spectral atlas of
Galactic standards has been created using high-resolution spectra of
O- and early B-type stars, drawn from both hemispheres and degraded
to the resolving power of the FLAMES spectra.

\section{Acknowledgements}
NM acknowledge the financial supported of the Bulgarian NSF (grant DO
02-85).  Based on observations from ESO programme 182.D-0222 (P.I.:
Evans).

\end{document}